\documentclass[aps, prd, preprint, groupedaddress, nofootinbib]{revtex4-1}

\usepackage{amsmath}
\usepackage{amssymb}
\usepackage{stmaryrd}
\usepackage{wasysym}
\usepackage{amsfonts}
\usepackage{graphicx}
\usepackage{array}
\usepackage{multirow}

\begin{document}



\title{Static Spherically Symmetric Solutions in Modified Entropic Gravity}






\author{
A. Rostami$^{1}$\footnote{abasat.rostami@aut.ac.ir},
M. Rostampour$^{1}$\footnote{mostafarostampour@aut.ac.ir},
A. Yarahmadi$^{2}$\footnote{amiryarahmadi@ipm.ir},
and
K. Rezazadeh$^{2}$\footnote{kazem.rezazadeh@ipm.ir}
}

\affiliation{
$^{1}$\small{Physics and Energy Engineering Department, Amirkabir University of Technology (Tehran Polytechnics), P.O. Box 159163-4311, Tehran, Iran}\\
$^{2}$\small{School of Astronomy, Institute for Research in Fundamental Sciences (IPM), P.O. Box 19395-5531, Tehran, Iran}
}

\date{\today}


\begin{abstract}

Within the entropic gravity paradigm introduced by Verlinde, one assumes that the microscopic degrees of freedom residing on the holographic screen obey the equipartition law of energy. Nevertheless, implications of statistical mechanics suggest that this energy sharing can acquire corrections that depend on temperature. Taking such modifications into account leads to altered gravitational field equations when derived from thermodynamic considerations. We solve the resulting modified Einstein equations in the case of a static, spherically symmetric spacetime and determine the general structure of the metric components. Our findings indicate that if the temperature correction function behaves as $f(T)\propto T^{2}$, the corresponding spacetime geometry reduces to the flat Minkowski metric. Therefore, to obtain small deviations from flatness, it is necessary to consider slight departures from this purely quadratic temperature dependence. Such deviations can be interpreted as encoding additional gravitational effects, potentially associated with matter contributions that are not entirely described by the holographic screen. As a result of these effects, the effective gravitational potential acquires a logarithmic correction, which can give rise to deviations from Newtonian gravity. This aspect is especially significant in the regime of very weak gravitational fields, where these corrections may affect particle motion and could carry implications for both astrophysical and cosmological contexts.

\end{abstract}



\maketitle



\section{Introduction}
\label{sec:int}

Over recent decades, extensive research has revealed deep connections between thermodynamics and gravity, significantly influencing our fundamental understanding of spacetime. The seminal contributions of Bekenstein and Hawking established that black holes can be described as thermodynamic systems: they are characterized by entropy and radiate thermally~\cite{Bardeen:1973gs, Hawking:1974rv, Hawking:1975vcx, Bekenstein:1973ur}. These results suggest that gravitational systems intrinsically exhibit thermodynamic behavior, supporting the perspective that gravity may not be a fundamental interaction, but rather an emergent phenomenon arising from underlying thermodynamic principles.

Another fundamental element in thermodynamic approaches to gravity is the Unruh effect~\cite{Unruh:1976db}, which establishes a direct connection between acceleration and temperature. According to this phenomenon, an observer with uniform acceleration in the quantum vacuum detects a thermal spectrum, with a temperature proportional to the magnitude of the acceleration~\cite{Unruh:1976db}. This striking result bridges concepts from acceleration, quantum field theory, and thermodynamics, and serves as a key ingredient in contemporary efforts to interpret gravity as an emergent phenomenon. It also offers a natural framework for incorporating modifications to gravity through background temperature effects.

A significant conceptual breakthrough was provided by Jacobson~\cite{Jacobson:1995ab}, who showed that the Einstein field equations can be viewed as equations of state governing a thermodynamic system. This perspective laid the foundation for Verlinde’s proposal of entropic gravity~\cite{Verlinde:2010hp}, where gravity arises as an entropic force associated with variations in the information content related to the positions of matter. By employing the holographic principle together with the equipartition law of energy on a holographic screen, Verlinde was able to recover both Newtonian gravity and Einstein’s general relativity from purely thermodynamic arguments~\cite{Verlinde:2010hp}.

The holographic principle, first introduced by 't~Hooft and later developed by Susskind~\cite{tHooft:1993dmi, Susskind:1994vu, Susskind:2005js}, asserts that all the physical information within a given volume of space can be entirely described by degrees of freedom defined on its boundary. This groundbreaking idea suggests that our three-dimensional universe may be understood as a holographic projection of information encoded on a two-dimensional surface, leading to a profound revision of our concepts of space, time, and information.

In Verlinde’s framework~\cite{Verlinde:2010hp}, it is assumed that the total energy associated with the holographic screen is uniformly distributed among its microscopic degrees of freedom. Accordingly, the energy carried by the bits of information on the holographic screen follows the equipartition law, $E = N k_B T/2$, where $N$ is the number of bits, $T$ is the temperature of the holographic screen, and $k_B$ denotes the Boltzmann constant.

In the present work, we adopt this perspective while allowing for temperature-dependent modifications to the equipartition law, which we express in the generalized form $E = N k_B T\, f(T)/2$, where $f(T)$ represents a temperature correction function. Insights from the statistical mechanics of condensed matter systems indicate that such corrections are essential, particularly at low temperatures, to account for deviations in heat capacity arising from the quantum statistical behavior of bosons~\cite{Reif1965, Pathria:1996hda}.

By incorporating thermal corrections to the equipartition law within Verlinde’s entropic force framework, one obtains a modified version of Einstein’s gravitational field equations~\cite{Sheykhi:2012vf, Rezazadeh:2025htb}. In this work, we analyze these modified equations in the case of a static, spherically symmetric spacetime and derive the corresponding form of the metric in the regime of extremely weak gravitational fields.

Our analysis indicates that in the limit of very low temperatures, if the temperature correction function $f(T)$ appearing in the modified equipartition law scales as $f(T)\propto T^{2}$, then the resulting static, spherically symmetric metric reduces to flat Minkowskian spacetime. Consequently, small departures from this behavior, which may be parametrized as $f(T)\propto T^{2+\epsilon}$ with $|\epsilon|\ll 1$, can give rise to slight deviations from a flat Minkowskian geometry.

Such small modifications of the spacetime metric naturally influence the motion of the test particles in the vicinity of the holographic screen. These effects become particularly relevant in the regime of extremely weak gravitational fields and may lead to observable consequences in both astrophysical and cosmological settings.

The structure of this paper is as follows. In Sec.~\ref{sec:Einstein}, we investigate temperature-dependent corrections to the energy equipartition law within the entropic force framework and explore their consequences for the Einstein gravitational field equations. Based on this formulation, in Sec.~\ref{sec:metric}, we solve the resulting modified field equations in a static, spherically symmetric spacetime, and derive the corresponding metric functions. Finally, in Sec.~\ref{sec:con}, we present our conclusions and outline possible directions for future work.


\section{Modified Einstein Equations}
\label{sec:Einstein}

We begin with the holographic principle, which states that the number of degrees of freedom $N$ associated with a holographic screen $\mathcal{S}$ is proportional to its area~\cite{Verlinde:2010hp},
\begin{equation}
N=\frac{A}{\ell_{P}^{2}} \, ,
\label{N}
\end{equation}
where $A$ denotes the area of the two-dimensional boundary and $\ell_{P}=\sqrt{\hbar G/c^{3}}$ is the Planck length. In the present work, we perform all calculations in natural units, setting $c = G = \hbar = k_{B} = 1$.

We then adopt a generalized form of the equipartition law for the total energy $E$ associated with the $N$ bits on the holographic screen with the temperature $T$,
\begin{equation}
E = \frac{1}{2} N T f(T) \, .
\label{E}
\end{equation}
In this equation, $f(T)$ denotes a temperature-dependent correction function that encodes deviations from the standard equipartition theorem due to quantum statistical effects. The temperature correction function $f(T)$ introduced in the above equation should be viewed as an effective phenomenological quantity in our analysis rather than a fundamental microscopic prediction. The philosophy underlying the present approach is similar to that of many generalized thermodynamic descriptions of gravity, in which one first explores the consequences of modified thermodynamic relations and subsequently investigates their microscopic origins.

Several generalized entropy formalisms have been proposed in recent years. Examples include Tsallis non-extensive entropy~\cite{Tsallis:1987eu}, Kaniadakis entropy~\cite{Kaniadakis:2002zz, Kaniadakis:2005zk, Kaniadakis:2009ka, Drepanou:2021jiv}, Barrow entropy~\cite{Barrow:2020tzx, Saridakis:2020zol}, Luciano-Saridakis~\cite{Luciano:2026ufu}, and other generalized entropy constructions. These frameworks modify the entropy-area relation and consequently alter the gravitational field equations obtained from thermodynamic arguments. The present work, however, differs conceptually from those approaches. Instead of modifying the entropy associated with the holographic screen, we introduce a correction at the level of the equipartition law itself. Consequently, the resulting modifications originate from the microscopic energy distribution among the holographic degrees of freedom rather than from a modification of the entropy functional.

Such corrections are physically motivated by statistical mechanics. In condensed matter systems, the classical equipartition theorem is known to fail at sufficiently low temperatures because quantum effects modify the distribution of energy among microscopic degrees of freedom. A well-known example is Debye theory~\cite{Reif1965, Pathria:1996hda}, where the internal energy exhibits nontrivial temperature dependence. Motivated by these considerations, we regard $f(T)$ as an effective function encoding possible deviations from classical equipartition on the holographic screen. A fundamental microscopic derivation of $f(T)$ from an underlying quantum theory of gravity remains an open problem and is left for future investigation.

Here, it is also convenient to introduce the dimensionless variable $x \equiv T_{\rm ref} / T$, where $T_{\rm ref}$ is a characteristic reference temperature of the system (for instance, the Debye temperature $T_D$ in specific realizations~\cite{Reif1965, Pathria:1996hda}), and $T$ represents the Unruh temperature associated with the holographic screen. The only physical requirement imposed on the function $f(T)$ is that it reproduces classical behavior in the high-temperature limit,
\begin{equation}
\lim_{T \to \infty} f(T) = \lim_{x \to 0} f(x) = 1\, .
\label{f-highT}
\end{equation}
The asymptotic form of $f(T)$ can be derived either from a theoretical approach, by specifying assumptions about the underlying energy distribution, or from a phenomenological perspective, by demanding consistency of the final results with observational data. In this work, as we will see in the next section, we derive the form of $f(T)$ in such a way that it can describe small deviations from the flat geometry of space-time in the presence of very weak gravitational fields.

By combining these considerations with the mass–energy equivalence relation $E=M$, we define the total thermodynamic mass $M_{\rm th}$ enclosed by the screen $\mathcal{S}$ as an integral over its surface,
\begin{equation}
M_{\rm th} = \int dM = \oint_\mathcal{S} \left( \frac{1}{2G\hbar} \, T f(T) \right) dA \, .
\label{Mth}
\end{equation}
To cast this expression in a covariant form, we identify the temperature $T$ with the Unruh temperature experienced by a static observer associated with a timelike Killing vector field $\xi^a$. This temperature is given by
\begin{equation}
T = \frac{\hbar a}{2\pi} \, ,
\label{T}
\end{equation}
where the proper acceleration $a$ is defined as $a = e^\phi N^b \nabla_b \phi$. Here, $\phi = \ln(-\xi^a\xi_a)/2$ denotes the redshift potential, and $N^b$ is the unit normal vector to the equipotential screen $\mathcal{S}$.

By substituting the covariant expression for the temperature, $T = \frac{\hbar}{2\pi} e^\phi N^b \nabla_b \phi$, we obtain the final form of the thermodynamic mass~\cite{Sheykhi:2012vf, Rezazadeh:2025htb},
\begin{equation}
M_{\rm th} = \oint_\mathcal{S} \left( \frac{1}{2G\hbar} \right)
\left( \frac{\hbar}{2\pi} e^\phi N^b \nabla_b \phi \right)
f(T)\, dA \, .
\label{Mth2}
\end{equation}
This relation provides the thermodynamic definition of the mass enclosed by the holographic screen, now generalized through the inclusion of the statistical correction function $f(T)$.

Next, we aim to construct a purely geometric notion of mass, denoted by $M_{\rm geo}$, which is identified with the thermodynamic mass $M_{\rm th}$. In standard general relativity, the Komar mass $M_K$ is defined as~\cite{Wald:1984rg}
\begin{equation}
M_K = -\frac{1}{8\pi G} \oint_\mathcal{S} \nabla^a \xi^b\, d\mathcal{S}_{ab} \, .
\label{MK}
\end{equation}
This expression can alternatively be written in the form of
\begin{equation}
M_K = \frac{1}{4\pi G} \oint_\mathcal{S} (e^\phi N^b \nabla_b \phi)\, dA \, .
\label{MK2}
\end{equation}

In the modified entropic gravity model, the function $f(T)$ appears in the modified energy equipartition law as a correction factor to the standard equipartition law. Given that, according to the equivalence principle, thermodynamic mass is equivalent to energy, it is therefore expected that this function also appears in the modified Komar mass relation. Hence, comparing Eq.~\eqref{MK2} with the thermodynamic mass $M_{\rm th}$ given in Eqs.~\eqref{Mth2}, we acquire
\begin{equation}
M_{\rm th} = \oint_\mathcal{S} f(T)\, dM_K \, .
\label{Mth-MK}
\end{equation}
This observation leads us to the central postulate of the present work: the appropriate geometric mass $M_{\rm geo}$ in this generalized entropic gravity framework is given by a modified Komar integral in which the correction function $f(T)$ is incorporated directly into the integrand,
\begin{equation}
M_{\rm geo} \equiv -\frac{1}{8\pi G} \oint_\mathcal{S} f(T)\, \nabla^a \xi^b\, d\mathcal{S}_{ab} \, .
\label{Mgeo}
\end{equation}
Since the temperature correction function $f(T)$ appears explicitly in the modified Komar mass relation, it is expected that this function and its derivative will appear in the final result, as we will see.

Now, we point out that the generalized Stokes' theorem states
\begin{equation}
\oint_S \omega = \int_\Sigma d\omega \, ,
\label{Stokes}
\end{equation}
to convert the surface integral over $\mathcal{S}$ into a volume integral over the enclosed hypersurface $\Sigma$. For an antisymmetric tensor field $B^{ab}$, this theorem yields~\cite{poisson2004relativist}
\begin{equation}
\oint_S B^{ab} d\mathcal{S}_{ab} = 2 \int_\Sigma \nabla_b B^{ab} d\Sigma_a \, .
\label{Stokes2}
\end{equation}
In our discussion, the term $B^{ab}$ is given by the following relation
\begin{equation}
B^{ab}=f(T)\nabla^{a}\xi^{b} \, .
\label{Bab}
\end{equation}
Using the Lorentz chain derivative rule, the covariant derivative of this term will be obtained as follows
\begin{equation}
\nabla_{b}\left[f(T)\nabla^{a}\xi^{b}\right]=\nabla_{b}f(T)\nabla^{a}\xi^{b}+f(T)\nabla_{b}\nabla^{a}\xi^{b} \, .
\label{CD-Bab}
\end{equation}
Thus, applying the the generalized Stokes' theorem, Eq.~\eqref{Stokes}, to $M_{\rm geo}$ in Eq.~\eqref{Mgeo}, we arrive at
\begin{equation}
M_{{\rm geo}}=-\frac{1}{8\pi G}\left[2\int_{\Sigma}\nabla_{b}\left(f(T)\nabla^{a}\xi^{b}\right)d\Sigma_{a}\right] \, .
\label{Mgeo2}
\end{equation}
Using the product rule for covariant differentiation, this becomes
\begin{equation}
M_{{\rm geo}}=-\frac{1}{4\pi G}\int_{\Sigma}\left[\nabla_{b}f(T)\left(\nabla^{a}\xi^{b}\right)+f(T)\left(\nabla_{b}\nabla^{a}\xi^{b}\right)\right]d\Sigma_{a} \, .
\label{Mgeo3}
\end{equation}
It is useful to note that the appearance of the covariant derivative of the function $f(T)$ in the thermodynamic mass relation is a consequence of the application of Stokes' theorem.

To simplify the second term, we use the standard identities for a Killing vector field $\xi^a$~\cite{Wald:1984rg},
\begin{align}
\nabla_b \nabla^a \xi^b &= -\nabla_b \nabla^b \xi^a \, ,
\\
\nabla_b \nabla^b \xi^a &= -R^a_{\ c} \xi^c \, ,
\end{align}
which together imply
\begin{equation}
\nabla_b \nabla^a \xi^b = R^a_{\ c} \xi^c \, .
\label{xi-R}
\end{equation}
Substituting Eq.~\eqref{xi-R} into Eq.~\eqref{Mgeo3}, we find
\begin{equation}
M_{{\rm geo}}=-\frac{1}{4\pi G}\int_{\Sigma}\left[\nabla_{b}f(T)\left(\nabla^{a}\xi^{b}\right)+f(T)R_{\ c}^{a}\xi^{c}\right]d\Sigma_{a} \, .
\label{Mgeo4}
\end{equation}
For a spacelike hypersurface $\Sigma$, the directed surface element is $d\Sigma_a = -n_a d\Sigma$, where $n^a$ is the future-directed unit normal vector. Therefore, the above expression can be rewritten as
\begin{equation}
M_{{\rm geo}}=\frac{1}{4\pi G}\int_{\Sigma}\left[f(T)R_{ac}\xi^{c}+\nabla_{a}f(T)\left(\nabla_{c}\xi^{c}\right)\right]n^{a}d\Sigma \, .
\label{Mgeo5}
\end{equation}
This expression provides a fully geometric representation of the total mass in terms of the generalized, statistically corrected spacetime structure.

The final step is to equate the geometric definition of mass $M_{\rm geo}$ with the matter-energy definition derived from the stress-energy tensor $\mathcal{T}_{ab}$. The corresponding matter mass $M_{\rm matter}$ is given by~\cite{Wald:1984rg}
\begin{equation}
M_{\rm matter} = 2 \int_\Sigma \left( \mathcal{T}_{ab} - \frac{1}{2} \mathcal{T} g_{ab} \right) n^a \xi^b d\Sigma \, .
\label{Mmatter}
\end{equation}

The fundamental postulate of gravity, $M_{\rm geo} = M_{\rm matter}$, then implies
\begin{equation}
\frac{1}{4\pi G} \int_\Sigma \left[ f(T) R_{ac} + g_{bc}\,\nabla_a \xi^c \nabla_c f(T) \right] n^a \xi^c d\Sigma
= 2 \int_\Sigma \left( \mathcal{T}_{ac} - \frac{1}{2} \mathcal{T} g_{ac} \right) n^a \xi^c d\Sigma \, ,
\label{Mgeo-Mmatter}
\end{equation}
where we have relabeled indices $b \to c$ for consistency. Since this relation must hold for any arbitrary hypersurface $\Sigma$, any timelike Killing vector $\xi^c$, and any unit normal $n^a$, the integrands themselves must be equal pointwise. This leads to the generalized Einstein field equations~\cite{Sheykhi:2012vf, Rezazadeh:2025htb}
\begin{equation}
f(T)\, R_{ab}
- e^{-2\phi}\,\xi_{b}\nabla_{a}\xi^{c}\nabla_{c} f(T)
= 8\pi G \left( \mathcal{T}_{ab} - \frac{1}{2}\mathcal{T} g_{ab} \right) \, .
\label{Einstein}
\end{equation}
This represents the most general form of the modified field equations emerging from the assumption that the holographic principle is subject to a non-classical statistical correction $f(T)$, which is not necessarily given by the Debye function. In the high-temperature limit corresponding to strong gravitational fields, one has $f(T)=1$, and the above equation reduces to the standard Einstein field equations in general relativity. In contrast, in the low-temperature regime associated with extremely weak gravitational fields, these thermal corrections can lead to significant deviations from general relativity, potentially modifying gravitational dynamics in astrophysical and cosmological contexts.

An important feature of Eq.~\eqref{Einstein} is the appearance of the non-standard term involving derivatives of the correction function $f(T)$. It is important to emphasize that this contribution is not introduced phenomenologically at the level of the field equations. Instead, it arises naturally from the generalized Komar construction.


\section{Solution for Static, Spherically Metric}
\label{sec:metric}

In this section, we focus on solving the modified Einstein equations~\eqref{Einstein} for a static, spherically symmetric spacetime described by the line element
\begin{equation}
ds^2 = -A(r)\, dt^2 + B(r)\, dr^2 + r^2 \left(d\theta^2 + \sin^2\theta \, d\phi^2\right) \, ,
\label{metric}
\end{equation}
where $A(r)$ and $B(r)$ are the metric coefficient functions, depending solely on the radial coordinate $r$. Our objective in what follows is to derive the explicit analytic expressions for these functions by solving the modified field equations.

The Christoffel symbols of the second kind are defined as
\begin{equation}
\Gamma_{\mu\nu}^{\lambda}
=\frac{1}{2}g^{\lambda\sigma}\left(\partial_{\mu}g_{\nu\sigma}
+\partial_{\nu}g_{\mu\sigma}
-\partial_{\sigma}g_{\mu\nu}\right) \, .
\label{Gamma}
\end{equation}
Using this definition, the non-vanishing Christoffel symbols for the metric~\eqref{metric} are obtained as
\begin{align}
\Gamma_{tr}^{t} &= \Gamma_{rt}^{t}=\frac{A'}{2A} \, ,
\label{Gammat}
\\
\Gamma_{tt}^{r} &= \frac{A'}{2B},\quad
\Gamma_{rr}^{r}=\frac{B'}{2B},\quad
\Gamma_{\theta\theta}^{r}=-\frac{r}{B},\quad
\Gamma_{\phi\phi}^{r}=-\frac{r\sin^{2}\theta}{B} \, ,
\label{Gammar}
\\
\Gamma_{r\theta}^{\theta} &= \Gamma_{\theta r}^{\theta}=\frac{1}{r},\quad
\Gamma_{\phi\phi}^{\theta}=-\sin\theta\cos\theta \, ,
\label{Gammatheta}
\\
\Gamma_{r\phi}^{\phi} &= \Gamma_{\phi r}^{\phi}=\frac{1}{r},\quad
\Gamma_{\theta\phi}^{\phi}=\Gamma_{\phi\theta}^{\phi}=\cot\theta \, .
\label{Gammaphi}
\end{align}
where the prime refers to the derivative with $r$. With these results, we can compute the Riemann tensor, defined by
\begin{equation}
R_{\sigma\mu\nu}^{\rho}
=\partial_{\mu}\Gamma_{\nu\sigma}^{\rho}
-\partial_{\nu}\Gamma_{\mu\sigma}^{\rho}
+\Gamma_{\mu\lambda}^{\rho}\Gamma_{\nu\sigma}^{\lambda}
-\Gamma_{\nu\lambda}^{\rho}\Gamma_{\mu\sigma}^{\lambda} \, .
\label{Reimann}
\end{equation}
The Ricci tensor follows from the contraction $R_{\mu\nu}\equiv R_{\mu\lambda\nu}^{\lambda}$. For the metric~\eqref{metric}, the non-zero components are
\begin{align}
R_{tt} &= \frac{A''}{2B}-\frac{A'B'}{4B^{2}}-\frac{A'^{2}}{4AB}+\frac{A'}{rB} \, ,
\label{Rtt}
\\
R_{rr} &= -\frac{A''}{2A}+\frac{A'B'}{4AB}+\frac{A'^{2}}{4A^{2}}+\frac{B'}{rB} \, ,
\label{Rrr}
\\
R_{\theta\theta} &= 1-\frac{rA'}{2AB}+\frac{rB'}{2B^{2}}-\frac{1}{B} \, ,
\label{Rthetatheta}
\\
R_{\phi\phi} &= \sin^{2}\theta\,R_{\theta\theta}
=\sin^{2}\theta\left(1-\frac{rA'}{2AB}+\frac{rB'}{2B^{2}}-\frac{1}{B}\right) \, .
\label{Rphiphi}
\end{align}
Finally, contracting the Ricci tensor yields the Ricci scalar $R\equiv R_{\mu}^{\mu}$,
\begin{equation}
R
=-\frac{A''}{AB}
+\frac{A'B'}{2AB^{2}}
+\frac{A'^{2}}{2A^{2}B}
+\frac{2B'}{rB^{2}}
-\frac{2A'}{rAB}
+\frac{2}{r^{2}}
-\frac{2}{r^{2}B} \, .
\label{R}
\end{equation}

We now consider vacuum solutions of the gravitational field equations, for which the stress-energy tensor vanishes,
\begin{equation}
\mathcal{T}^{\mu}_{\nu} = 0 \, .
\label{Tmunu0}
\end{equation}
Consequently, the modified Einstein equations~\eqref{Einstein} reduce to
\begin{equation}
f(T)R_{\mu\nu}
- e^{-2\phi}\,\xi_{\nu}\nabla_{\mu}\xi^{\lambda}\nabla_{\lambda} f(T)
= 0 \, .
\label{Einstein-Tmunu0}
\end{equation}
Here, it is essential to clarify the meaning of vacuum within the present framework. Throughout this work, the condition~\eqref{Tmunu0} denotes the absence of ordinary matter fields. However, unlike standard General Relativity, the modified field equations still contain non-trivial geometric contributions generated by the temperature-dependent correction function $f(T)$. Consequently, the resulting spacetime is not necessarily Ricci-flat even in the absence of ordinary matter. Equation~\eqref{Einstein-Tmunu0} may therefore be rewritten schematically as
\begin{equation}
R_{\mu \nu}=\mathcal{T}^{\rm eff}_{\mu \nu} \, ,
\label{Rmunu-Teffmunu}
\end{equation}
where the effective source tensor
\begin{equation}
\mathcal{T}^{\rm eff}_{\mu \nu} = \frac{1}{f(T)} e^{-2\phi} \xi_\nu\nabla_\mu\xi^\lambda\nabla_\lambda f(T) \, ,
\label{Teffmunu}
\end{equation}
encodes the thermodynamic corrections associated with the modified equipartition law. Consequently, the solutions presented in this work should be interpreted as matter-vacuum solutions rather than vacuum solutions in the standard General Relativity sense. Only when both conditions $\mathcal{T}^{\mu}_{\nu} = 0$ and $\nabla_\mu f(T) = 0$ are simultaneously satisfied do the modified equations reduce to the ordinary vacuum Einstein equations.

Having fully specified the spacetime geometry together with the explicit form of the Christoffel symbols, we are now in a position to determine the Killing vectors associated with the spacetime. To this end, we seek solutions to the Killing equation,
\begin{equation}
\nabla_\mu \xi_\nu + \nabla_\nu \xi_\mu = 0 \, ,
\end{equation}
which encodes the continuous symmetries of the metric.

At this point, we make use of the definition of the covariant derivative,
\begin{equation}
\nabla_\mu \xi_\nu = \partial_\mu \xi_\nu - \Gamma^\lambda_{\mu\nu} \xi_\lambda \, .
\label{covDxi}
\end{equation}
Accordingly, the Killing equation can be equivalently expressed as
\begin{equation}
\partial_\mu \xi_\nu + \partial_\nu \xi_\mu - 2 \Gamma^\lambda_{\mu\nu} \xi_\lambda = 0 \, .
\label{pDxi}
\end{equation}

The covariant components of the Killing vector are obtained by lowering the index with the metric,
\begin{equation}
\xi_{\mu}=g_{\mu\nu}\xi^{\nu}
=\left(-A \, \xi^{t},\,B \, \xi^{r},\,r^{2}\xi^{\theta},\,r^{2}\sin^{2}\theta\,\xi^{\phi}\right) \, .
\label{covxi}
\end{equation}

One of the most important symmetries that the metric \eqref{metric} follows is symmetry under time translation, because this metric is completely independent of the time coordinate $t$. Therefore, the vector $\xi^\mu = (1, 0, 0, 0)$ satisfies the Killing equations~\eqref{pDxi} identically, confirming that it generates a symmetry under time translations. With this Killing vector, the modified Einstein equations~\eqref{Einstein-Tmunu0} lead to
\begin{align}
& f(r)\left(\frac{A''}{2B}-\frac{A'B'}{4B^{2}}-\frac{A'^{2}}{4AB}+\frac{A'}{rB}\right)+\frac{A'f'(r)}{2B}=0 \, ,
\label{Einstein00}
\\
& f(r)\left(-\frac{A''}{2A}+\frac{A'B'}{4AB}+\frac{A'^{2}}{4A^{2}}+\frac{B'}{rB}\right)=0 \, ,
\label{Einstein11}
\\
& f(r)\left(1-\frac{rA'}{2AB}+\frac{rB'}{2B^{2}}-\frac{1}{B}\right)=0 \, .
\label{Einstein22}
\end{align}
By combining Eqs.~\eqref{Einstein00} and \eqref{Einstein11} to eliminate $A''$, we obtain
\begin{equation}
BA'\left(rf'+2f\right)+2AfB'=0 \, .
\label{Einstein1}
\end{equation}
Moreover, Eq.~\eqref{Einstein22} can be straightforwardly rewritten as
\begin{equation}
1-\frac{rA'}{2AB}+\frac{rB'}{2B^{2}}-\frac{1}{B}=0 \, .
\label{Einstein2}
\end{equation}
Equations~\eqref{Einstein1} and \eqref{Einstein2} constitute the two key relations that determine the metric functions $A(r)$ and $B(r)$ in Eq.~\eqref{metric}.

In what follows, we derive the relation for the gravitational acceleration within our framework. For this purpose, we begin with the geodesic equations, which in general take the form
\begin{equation}
\frac{\partial^{2}x^{\mu}}{\partial\tau^{2}}+\text{\ensuremath{\Gamma}}_{\nu\lambda}^{\mu}\frac{\partial x^{\text{\ensuremath{\lambda}}}}{\partial\tau}\frac{\partial x^{\text{\ensuremath{\nu}}}}{\partial\tau}=0 \, .
\label{geodesic}
\end{equation}
For the metric~\eqref{metric}, this equation yields the following set of component equations
\begin{align}
& \frac{\partial^{2}t}{\partial\tau^{2}}+\frac{A'}{A}\frac{\partial t}{\partial\tau}\frac{\partial r}{\partial\tau}=0 \, ,
\label{geodesic0}
\\
& \frac{\partial^{2}r}{\partial\tau^{2}}+\frac{A'}{2B}\left(\frac{\partial t}{\partial\tau}\right)^{2}+\frac{B'}{2B}\left(\frac{\partial r}{\partial\tau}\right)^{2}-\frac{r}{B}\left(\frac{\partial\theta}{\partial\tau}\right)^{2}-\frac{r\sin^{2}\theta}{B}\left(\frac{\partial\phi}{\partial\tau}\right)^{2}=0 \, ,
\label{geodesic1}
\\
& \frac{\partial^{2}\theta}{\partial\tau^{2}}-\sin\theta\cos\theta\left(\frac{\partial\phi}{\partial\tau}\right)^{2}+\frac{2}{r}\left(\frac{\partial r}{\partial\tau}\right)\left(\frac{\partial\theta}{\partial\tau}\right)=0 \, ,
\label{geodesic2}
\\
& \frac{\partial^{2}\phi}{\partial\tau^{2}}+2\cot\theta\left(\frac{\partial\theta}{\partial\tau}\right)\left(\frac{\partial\phi}{\partial\tau}\right)+\frac{2}{r}\left(\frac{\partial r}{\partial\tau}\right)\left(\frac{\partial\phi}{\partial\tau}\right)=0 \, .
\label{geodesic3}
\end{align}

In our analysis, we consider the regime of weak gravitational fields and particle velocities that are much smaller than the speed of light. Under these assumptions, the motion can be treated in the non-relativistic limit. Consequently, in Eq.~\eqref{geodesic0}, the second term on the left-hand side is negligible compared to the first term~\cite{weinberg2013gravitation}. Similarly, in Eq.~\eqref{geodesic1}, only the first two terms contribute significantly, while the remaining terms can be safely neglected~\cite{weinberg2013gravitation}. Therefore, Eqs.~\eqref{geodesic0} and \eqref{geodesic1} reduce approximately to
\begin{align}
& \frac{\partial^{2}t}{\partial\tau^{2}}\approx0 \, ,
\label{geodesic0approx}
\\
& \frac{\partial^{2}r}{\partial\tau^{2}}+\frac{A'}{2B}\left(\frac{\partial t}{\partial\tau}\right)^{2}\approx0 \, .
\label{geodesic1approx}
\end{align}
From these relations, the gravitational acceleration in our model can be expressed as
\begin{equation}
a=\frac{\partial^{2}r}{\partial t^{2}}\approx-\frac{A'}{2B} \, .
\label{a}
\end{equation}
This expression provides the general form of the gravitational acceleration for the static, spherically symmetric metric~\eqref{metric} in the weak-field and low-velocity limit.

Now, we point out that from Eqs.~\eqref{Einstein1} and \eqref{Einstein2}, it is evident that choosing the correction function in the form $f(r)=f_{0}/r^{2}$, where $f_0$ is a constant parameter, leads to the simple solution $A(r)=B(r)=1$ that satisfies the equations. This corresponds to a flat Minkowskian spacetime and describes the situation in which a test particle near the holographic screen experiences negligible gravitational influence from the enclosed mass distribution, as well as from external sources. To incorporate the effects of these mass distributions in the weak-field regime, we introduce a small deviation from the scaling $f(r)=f_{0}/r^{2}$ and generalize it as
\begin{equation}
f(r)=\frac{f_{0}}{r^{2+\delta}} \, .
\label{f-r}
\end{equation}
Here, $\delta$ is a constant parameter satisfying $|\delta|\ll 1$, ensuring that the deviation from the exact scaling remains small.

Substituting Eq.~\eqref{f-r} into Eq.~\eqref{Einstein1} yields
\begin{equation}
\delta\,B(r)A'(r)-2A(r)B'(r)=0 \, .
\label{Einstein1-f-r}
\end{equation}
The solution of this differential equation expresses $B(r)$ in terms of $A(r)$ as
\begin{equation}
B(r)=A^{\delta/2}(r) \, ,
\label{B-A}
\end{equation}
where the integration constant has been fixed to unity to ensure asymptotic flatness. Substituting this relation back into Eq.~\eqref{Einstein2}, we find
\begin{equation}
\frac{1}{4}(\delta-2)rA^{-(\delta+2)/2}(r)A'(r)-A^{-\delta/2}(r)+1=0 \, .
\label{Einstein2-f-r}
\end{equation}
This differential equation can then be solved to determine the explicit radial dependence of $A(r)$. The resulting metric functions take the following forms
\begin{align}
A(r) &= \left(1+C_{A}r^{\frac{2\delta}{2-\delta}}\right)^{-\frac{2}{\delta}} \, ,
\label{A}
\\
B(r) &= \frac{1}{1+C_{A}r^{\frac{2\delta}{2-\delta}}} \, ,
\label{B}
\end{align}
where $C_{A}$ is an integration constant. Since the spacetime under consideration is assumed to deviate only slightly from flat space, the second term inside the parentheses in Eq.~\eqref{A} is much smaller than fist term.

By substituting Eqs.~\eqref{A} and \eqref{B} into Eq.~\eqref{a}, we obtain the relation between the magnitude of the centripetal acceleration and the radial coordinate as
\begin{equation}
\left|a(r)\right|=\frac{2}{2-\delta}\left|C_{A}\right|r^{-\frac{2-3\delta}{2-\delta}}\left(1+C_{A}r^{\frac{2\delta}{2-\delta}}\right)^{-\frac{2}{\delta}}\approx\frac{2}{2-\delta}\left|C_{A}\right|r^{-\frac{2-3\delta}{2-\delta}} \, .
\label{a-r}
\end{equation}
Inserting this into the Unruh temperature relation~\eqref{T}, the temperature associated with the holographic screen can be written as
\begin{equation}
T\approx\frac{1}{(2-\delta)\pi}\left|C_{A}\right|r^{-\frac{2-3\delta}{2-\delta}} \, .
\label{T-r}
\end{equation}

We can now express the temperature correction function $f(T)$ purely in terms of $T$. To this end, we eliminate the radial coordinate $r$ between Eqs.~\eqref{f-r} and \eqref{T-r}, which leads to
\begin{equation}
f(T)=\alpha \, T^{2+\epsilon} \, ,
\label{f-T}
\end{equation}
where the constants $\alpha$ and $\epsilon$ are defined as
\begin{align}
\alpha &= \left(\frac{(2-\delta)\pi}{\left|C_{A}\right|}\right)^{2+\epsilon} f_{0} \approx \frac{4 \pi^2}{\left|C_{A}\right|^2} \, f_0 \, ,
\label{alpha}
\\
\epsilon &= \frac{(6-\delta)\delta}{2-3\delta}\approx 3 \delta \, .
\label{epsilon}
\end{align}
As it is obvious from Eq.~\eqref{epsilon}, the parameter $\epsilon$ is of the same order as $\delta$ and therefore remains much smaller than unity. In the limit $\epsilon \to 0$, the temperature dependence in Eq.~\eqref{f-T} reduces to $f(T)\propto T^{2}$, corresponding to the flat Minkowskian spacetime limit.

In this work, we have determined the functional form of the temperature correction $f(T)$ in a manner that allows it to effectively capture gravitational effects in the weak-field regime. In other words, Eq.~\eqref{f-T} specifies the appropriate form of the temperature correction required to incorporate small deviations from the flat Minkowski spacetime within the modified entropic gravity framework. In particular, the expression given in Eq.~\eqref{f-T} enables the study of subtle gravitational effects arising from both the mass distribution enclosed by the holographic screen and the influence of external masses, all within the limit of weak gravitational fields. This result may have important implications for astrophysical and cosmological applications and can be further explored in future investigations.

The modified solutions obtained in this section allow us to determine the effective gravitational potential associated with the weak-field regime. Starting from Eq.~\eqref{a-r}, and using the standard weak-field relation
\begin{equation}
a(r)=-\frac{d\Phi(r)}{dr} \, ,
\label{a-Phi}
\end{equation}
we acquire
\begin{equation}
\Phi(r)= \frac{2|C_A|}{2-\delta} \int r^{-\frac{2-3\delta}{2-\delta}} dr \, .
\label{Phi-r-int}
\end{equation}
Performing the integration yields
\begin{equation}
\Phi(r)= -\frac{|C_A|}{\delta} \, r^{\frac{2\delta}{2-\delta}} +\Phi_0 \, ,
\label{Phi-r}
\end{equation}
where $\Phi_0$ is an integration constant. For sufficiently small values of $\delta$, one may use
\begin{equation}
r^{\frac{2\delta}{2-\delta}}= \exp\left( \frac{2\delta}{2-\delta}\ln r \right) \approx 1+\delta \, \ln r \, ,
\label{ln_r}
\end{equation}
which leads to
\begin{equation}
\Phi(r) \approx -|C_A|\ln r+\mathrm{constant} \, .
\label{Phi-r-small_delta}
\end{equation}
Thus, small departures from the exact scaling $f(T)\propto T^2$ generate logarithmic corrections to the effective gravitational potential. Such corrections become increasingly important at large distances and may modify gravitational dynamics in the extremely weak-field regime. As discussed in~\cite{Anand:2025rjg, Anand:2025cer}, several modified gravity theories predict deviations from the Newtonian gravitational potential in the spacetime surrounding black holes, potentially leading to observable phenomenological consequences. A detailed investigation of the implications of the logarithmic corrections derived in our setting, including their effects on black hole physics, galactic dynamics, and large-scale structure formation, is beyond the scope of the present work and will be presented elsewhere.


\section{Conclusions}
\label{sec:con}

In this work, we have investigated a generalized entropic gravity framework by introducing temperature-dependent corrections to the equipartition law of energy defined on the holographic screen. Motivated by ideas from statistical mechanics, we proposed a correction function $f(T)$ that alters the standard distribution of energy among the microscopic degrees of freedom and analyzed its consequences for the resulting gravitational dynamics.

By matching the thermodynamic definition of mass with a generalized geometric (Komar-like) mass, we obtained modified Einstein field equations that incorporate the effects of the temperature correction function $f(T)$. Importantly, these equations reduce to the standard Einstein field equations in the high-temperature limit, where $f(T) = 1$, thereby ensuring consistency with general relativity in the strong-field regime.

Restricting attention to a static, spherically symmetric spacetime, we solved the modified field equations in vacuum and derived explicit expressions for the metric functions. Our results show that when the correction function scales as $f(T)\propto T^{2}$ in the low-temperature regime, the spacetime reduces exactly to the flat Minkowskian form. This behavior implies that this particular temperature dependence effectively eliminates gravitational effects in the limit of extremely weak fields.

To incorporate small but non-zero gravitational effects, we introduced a slight deviation from this scaling characterized by a small parameter $\epsilon$. The resulting framework offers a consistent description of subtle gravitational phenomena originating both from the mass enclosed by the holographic screen and from additional contributions associated with external matter distributions. These modifications give rise to a non-trivial spacetime geometry with weak curvature and, correspondingly, alter the radial behavior of the gravitational acceleration. Moreover, these modifications induce a logarithmic correction to the effective gravitational potential, which may lead to observable deviations from Newtonian gravity in the weak-field regime.

Our results indicate that temperature-dependent corrections to the equipartition law may have a non-negligible impact on gravitational dynamics in the infrared regime. This framework provides a new perspective for exploring weak-field gravity and may yield valuable insights into astrophysical and cosmological settings, especially in situations where deviations from standard Newtonian or relativistic behavior are present.

Future investigations could focus on confronting this model with observational data, as well as studying its implications for back hole physics, galactic dynamics, and the formation of large-scale structure. It would also be of interest to extend the present analysis to more general spacetimes and time-dependent settings. Furthermore, gaining a deeper understanding of the microscopic origin of the correction function $f(T)$ from a fundamental quantum theory of gravity would be highly desirable.


\section*{Acknowledgements}

The authors are grateful to the referee for his/her valuable and insightful comments, which have helped improve the clarity and quality of the manuscript.


%
%


\bibliography{static_spherical_solution}

\begin{thebibliography}{27}%
\makeatletter
\providecommand \@ifxundefined [1]{%
 \@ifx{#1\undefined}
}%
\providecommand \@ifnum [1]{%
 \ifnum #1\expandafter \@firstoftwo
 \else \expandafter \@secondoftwo
 \fi
}%
\providecommand \@ifx [1]{%
 \ifx #1\expandafter \@firstoftwo
 \else \expandafter \@secondoftwo
 \fi
}%
\providecommand \natexlab [1]{#1}%
\providecommand \enquote  [1]{``#1''}%
\providecommand \bibnamefont  [1]{#1}%
\providecommand \bibfnamefont [1]{#1}%
\providecommand \citenamefont [1]{#1}%
\providecommand \href@noop [0]{\@secondoftwo}%
\providecommand \href [0]{\begingroup \@sanitize@url \@href}%
\providecommand \@href[1]{\@@startlink{#1}\@@href}%
\providecommand \@@href[1]{\endgroup#1\@@endlink}%
\providecommand \@sanitize@url [0]{\catcode `\\12\catcode `\$12\catcode
  `\&12\catcode `\#12\catcode `\^12\catcode `\_12\catcode `\%12\relax}%
\providecommand \@@startlink[1]{}%
\providecommand \@@endlink[0]{}%
\providecommand \url  [0]{\begingroup\@sanitize@url \@url }%
\providecommand \@url [1]{\endgroup\@href {#1}{\urlprefix }}%
\providecommand \urlprefix  [0]{URL }%
\providecommand \Eprint [0]{\href }%
\providecommand \doibase [0]{http://dx.doi.org/}%
\providecommand \selectlanguage [0]{\@gobble}%
\providecommand \bibinfo  [0]{\@secondoftwo}%
\providecommand \bibfield  [0]{\@secondoftwo}%
\providecommand \translation [1]{[#1]}%
\providecommand \BibitemOpen [0]{}%
\providecommand \bibitemStop [0]{}%
\providecommand \bibitemNoStop [0]{.\EOS\space}%
\providecommand \EOS [0]{\spacefactor3000\relax}%
\providecommand \BibitemShut  [1]{\csname bibitem#1\endcsname}%
\let\auto@bib@innerbib\@empty
\bibitem [{\citenamefont {Bardeen}\ \emph {et~al.}(1973)\citenamefont
  {Bardeen}, \citenamefont {Carter},\ and\ \citenamefont
  {Hawking}}]{Bardeen:1973gs}%
  \BibitemOpen
  \bibfield  {author} {\bibinfo {author} {\bibfnamefont {J.~M.}\ \bibnamefont
  {Bardeen}}, \bibinfo {author} {\bibfnamefont {B.}~\bibnamefont {Carter}}, \
  and\ \bibinfo {author} {\bibfnamefont {S.~W.}\ \bibnamefont {Hawking}},\
  }\href {\doibase 10.1007/BF01645742} {\bibfield  {journal} {\bibinfo
  {journal} {Commun. Math. Phys.}\ }\textbf {\bibinfo {volume} {31}},\ \bibinfo
  {pages} {161} (\bibinfo {year} {1973})}\BibitemShut {NoStop}%
\bibitem [{\citenamefont {Hawking}(1974)}]{Hawking:1974rv}%
  \BibitemOpen
  \bibfield  {author} {\bibinfo {author} {\bibfnamefont {S.~W.}\ \bibnamefont
  {Hawking}},\ }\href {\doibase 10.1038/248030a0} {\bibfield  {journal}
  {\bibinfo  {journal} {Nature}\ }\textbf {\bibinfo {volume} {248}},\ \bibinfo
  {pages} {30} (\bibinfo {year} {1974})}\BibitemShut {NoStop}%
\bibitem [{\citenamefont {Hawking}(1975)}]{Hawking:1975vcx}%
  \BibitemOpen
  \bibfield  {author} {\bibinfo {author} {\bibfnamefont {S.~W.}\ \bibnamefont
  {Hawking}},\ }\href {\doibase 10.1007/BF02345020} {\bibfield  {journal}
  {\bibinfo  {journal} {Commun. Math. Phys.}\ }\textbf {\bibinfo {volume}
  {43}},\ \bibinfo {pages} {199} (\bibinfo {year} {1975})},\ \bibinfo {note}
  {[Erratum: Commun.Math.Phys. 46, 206 (1976)]}\BibitemShut {NoStop}%
\bibitem [{\citenamefont {Bekenstein}(1973)}]{Bekenstein:1973ur}%
  \BibitemOpen
  \bibfield  {author} {\bibinfo {author} {\bibfnamefont {J.~D.}\ \bibnamefont
  {Bekenstein}},\ }\href {\doibase 10.1103/PhysRevD.7.2333} {\bibfield
  {journal} {\bibinfo  {journal} {Phys. Rev. D}\ }\textbf {\bibinfo {volume}
  {7}},\ \bibinfo {pages} {2333} (\bibinfo {year} {1973})}\BibitemShut
  {NoStop}%
\bibitem [{\citenamefont {Unruh}(1976)}]{Unruh:1976db}%
  \BibitemOpen
  \bibfield  {author} {\bibinfo {author} {\bibfnamefont {W.~G.}\ \bibnamefont
  {Unruh}},\ }\href {\doibase 10.1103/PhysRevD.14.870} {\bibfield  {journal}
  {\bibinfo  {journal} {Phys. Rev. D}\ }\textbf {\bibinfo {volume} {14}},\
  \bibinfo {pages} {870} (\bibinfo {year} {1976})}\BibitemShut {NoStop}%
\bibitem [{\citenamefont {Jacobson}(1995)}]{Jacobson:1995ab}%
  \BibitemOpen
  \bibfield  {author} {\bibinfo {author} {\bibfnamefont {T.}~\bibnamefont
  {Jacobson}},\ }\href {\doibase 10.1103/PhysRevLett.75.1260} {\bibfield
  {journal} {\bibinfo  {journal} {Phys. Rev. Lett.}\ }\textbf {\bibinfo
  {volume} {75}},\ \bibinfo {pages} {1260} (\bibinfo {year} {1995})},\ \Eprint
  {http://arxiv.org/abs/gr-qc/9504004} {arXiv:gr-qc/9504004} \BibitemShut
  {NoStop}%
\bibitem [{\citenamefont {Verlinde}(2011)}]{Verlinde:2010hp}%
  \BibitemOpen
  \bibfield  {author} {\bibinfo {author} {\bibfnamefont {E.~P.}\ \bibnamefont
  {Verlinde}},\ }\href {\doibase 10.1007/JHEP04(2011)029} {\bibfield  {journal}
  {\bibinfo  {journal} {JHEP}\ }\textbf {\bibinfo {volume} {04}},\ \bibinfo
  {pages} {029} (\bibinfo {year} {2011})},\ \Eprint
  {http://arxiv.org/abs/1001.0785} {arXiv:1001.0785 [hep-th]} \BibitemShut
  {NoStop}%
\bibitem [{\citenamefont {'t~Hooft}(1993)}]{tHooft:1993dmi}%
  \BibitemOpen
  \bibfield  {author} {\bibinfo {author} {\bibfnamefont {G.}~\bibnamefont
  {'t~Hooft}},\ }\href@noop {} {\bibfield  {journal} {\bibinfo  {journal}
  {Conf. Proc. C}\ }\textbf {\bibinfo {volume} {930308}},\ \bibinfo {pages}
  {284} (\bibinfo {year} {1993})},\ \Eprint
  {http://arxiv.org/abs/gr-qc/9310026} {arXiv:gr-qc/9310026} \BibitemShut
  {NoStop}%
\bibitem [{\citenamefont {Susskind}(1995)}]{Susskind:1994vu}%
  \BibitemOpen
  \bibfield  {author} {\bibinfo {author} {\bibfnamefont {L.}~\bibnamefont
  {Susskind}},\ }\href {\doibase 10.1063/1.531249} {\bibfield  {journal}
  {\bibinfo  {journal} {J. Math. Phys.}\ }\textbf {\bibinfo {volume} {36}},\
  \bibinfo {pages} {6377} (\bibinfo {year} {1995})},\ \Eprint
  {http://arxiv.org/abs/hep-th/9409089} {arXiv:hep-th/9409089} \BibitemShut
  {NoStop}%
\bibitem [{\citenamefont {Susskind}\ and\ \citenamefont
  {Lindesay}(2005)}]{Susskind:2005js}%
  \BibitemOpen
  \bibfield  {author} {\bibinfo {author} {\bibfnamefont {L.}~\bibnamefont
  {Susskind}}\ and\ \bibinfo {author} {\bibfnamefont {J.}~\bibnamefont
  {Lindesay}},\ }\href@noop {} {\emph {\bibinfo {title} {An introduction to
  black holes, information and the string theory revolution: The holographic
  universe}}}\ (\bibinfo  {publisher} {World Scientific},\ \bibinfo {year}
  {2005})\BibitemShut {NoStop}%
\bibitem [{\citenamefont {Reif}(1965)}]{Reif1965}%
  \BibitemOpen
  \bibfield  {author} {\bibinfo {author} {\bibfnamefont {F.}~\bibnamefont
  {Reif}},\ }\href@noop {} {\emph {\bibinfo {title} {Fundamentals of
  Statistical and Thermal Physics}}}\ (\bibinfo  {publisher} {McGraw Hill},\
  \bibinfo {address} {Tokyo},\ \bibinfo {year} {1965})\BibitemShut {NoStop}%
\bibitem [{\citenamefont {Pathria}(1996)}]{Pathria:1996hda}%
  \BibitemOpen
  \bibfield  {author} {\bibinfo {author} {\bibfnamefont {R.~K.}\ \bibnamefont
  {Pathria}},\ }\href@noop {} {\emph {\bibinfo {title} {{Statistical
  Mechanics}}}},\ \bibinfo {edition} {2nd}\ ed.\ (\bibinfo  {publisher}
  {Butterworth-Heinemann},\ \bibinfo {year} {1996})\BibitemShut {NoStop}%
\bibitem [{\citenamefont {Sheykhi}\ and\ \citenamefont
  {Rezazadeh~Sarab}(2012)}]{Sheykhi:2012vf}%
  \BibitemOpen
  \bibfield  {author} {\bibinfo {author} {\bibfnamefont {A.}~\bibnamefont
  {Sheykhi}}\ and\ \bibinfo {author} {\bibfnamefont {K.}~\bibnamefont
  {Rezazadeh~Sarab}},\ }\href {\doibase 10.1088/1475-7516/2012/10/012}
  {\bibfield  {journal} {\bibinfo  {journal} {JCAP}\ }\textbf {\bibinfo
  {volume} {10}},\ \bibinfo {pages} {012} (\bibinfo {year} {2012})},\ \Eprint
  {http://arxiv.org/abs/1206.1030} {arXiv:1206.1030 [physics.gen-ph]}
  \BibitemShut {NoStop}%
\bibitem [{\citenamefont {Rezazadeh}(2025)}]{Rezazadeh:2025htb}%
  \BibitemOpen
  \bibfield  {author} {\bibinfo {author} {\bibfnamefont {K.}~\bibnamefont
  {Rezazadeh}},\ }\href@noop {} {\  (\bibinfo {year} {2025})},\ \Eprint
  {http://arxiv.org/abs/2503.08236} {arXiv:2503.08236 [gr-qc]} \BibitemShut
  {NoStop}%
\bibitem [{\citenamefont {Tsallis}(1988)}]{Tsallis:1987eu}%
  \BibitemOpen
  \bibfield  {author} {\bibinfo {author} {\bibfnamefont {C.}~\bibnamefont
  {Tsallis}},\ }\href {\doibase 10.1007/BF01016429} {\bibfield  {journal}
  {\bibinfo  {journal} {J. Statist. Phys.}\ }\textbf {\bibinfo {volume} {52}},\
  \bibinfo {pages} {479} (\bibinfo {year} {1988})}\BibitemShut {NoStop}%
\bibitem [{\citenamefont {Kaniadakis}(2002)}]{Kaniadakis:2002zz}%
  \BibitemOpen
  \bibfield  {author} {\bibinfo {author} {\bibfnamefont {G.}~\bibnamefont
  {Kaniadakis}},\ }\href {\doibase 10.1103/PhysRevE.66.056125} {\bibfield
  {journal} {\bibinfo  {journal} {Phys. Rev. E}\ }\textbf {\bibinfo {volume}
  {66}},\ \bibinfo {pages} {056125} (\bibinfo {year} {2002})},\ \Eprint
  {http://arxiv.org/abs/cond-mat/0210467} {arXiv:cond-mat/0210467} \BibitemShut
  {NoStop}%
\bibitem [{\citenamefont {Kaniadakis}(2005)}]{Kaniadakis:2005zk}%
  \BibitemOpen
  \bibfield  {author} {\bibinfo {author} {\bibfnamefont {G.}~\bibnamefont
  {Kaniadakis}},\ }\href {\doibase 10.1103/PhysRevE.72.036108} {\bibfield
  {journal} {\bibinfo  {journal} {Phys. Rev. E}\ }\textbf {\bibinfo {volume}
  {72}},\ \bibinfo {pages} {036108} (\bibinfo {year} {2005})},\ \Eprint
  {http://arxiv.org/abs/cond-mat/0507311} {arXiv:cond-mat/0507311} \BibitemShut
  {NoStop}%
\bibitem [{\citenamefont {Kaniadakis}(2009)}]{Kaniadakis:2009ka}%
  \BibitemOpen
  \bibfield  {author} {\bibinfo {author} {\bibfnamefont {G.}~\bibnamefont
  {Kaniadakis}},\ }\href {\doibase 10.1140/epjb/e2009-00161-0} {\bibfield
  {journal} {\bibinfo  {journal} {Eur. Phys. J. B}\ }\textbf {\bibinfo {volume}
  {70}},\ \bibinfo {pages} {3} (\bibinfo {year} {2009})},\ \Eprint
  {http://arxiv.org/abs/0904.4180} {arXiv:0904.4180 [cond-mat.stat-mech]}
  \BibitemShut {NoStop}%
\bibitem [{\citenamefont {Drepanou}\ \emph {et~al.}(2022)\citenamefont
  {Drepanou}, \citenamefont {Lymperis}, \citenamefont {Saridakis},\ and\
  \citenamefont {Yesmakhanova}}]{Drepanou:2021jiv}%
  \BibitemOpen
  \bibfield  {author} {\bibinfo {author} {\bibfnamefont {N.}~\bibnamefont
  {Drepanou}}, \bibinfo {author} {\bibfnamefont {A.}~\bibnamefont {Lymperis}},
  \bibinfo {author} {\bibfnamefont {E.~N.}\ \bibnamefont {Saridakis}}, \ and\
  \bibinfo {author} {\bibfnamefont {K.}~\bibnamefont {Yesmakhanova}},\ }\href
  {\doibase 10.1140/epjc/s10052-022-10415-9} {\bibfield  {journal} {\bibinfo
  {journal} {Eur. Phys. J. C}\ }\textbf {\bibinfo {volume} {82}},\ \bibinfo
  {pages} {449} (\bibinfo {year} {2022})},\ \Eprint
  {http://arxiv.org/abs/2109.09181} {arXiv:2109.09181 [gr-qc]} \BibitemShut
  {NoStop}%
\bibitem [{\citenamefont {Barrow}(2020)}]{Barrow:2020tzx}%
  \BibitemOpen
  \bibfield  {author} {\bibinfo {author} {\bibfnamefont {J.~D.}\ \bibnamefont
  {Barrow}},\ }\href {\doibase 10.1016/j.physletb.2020.135643} {\bibfield
  {journal} {\bibinfo  {journal} {Phys. Lett. B}\ }\textbf {\bibinfo {volume}
  {808}},\ \bibinfo {pages} {135643} (\bibinfo {year} {2020})},\ \Eprint
  {http://arxiv.org/abs/2004.09444} {arXiv:2004.09444 [gr-qc]} \BibitemShut
  {NoStop}%
\bibitem [{\citenamefont {Saridakis}(2020)}]{Saridakis:2020zol}%
  \BibitemOpen
  \bibfield  {author} {\bibinfo {author} {\bibfnamefont {E.~N.}\ \bibnamefont
  {Saridakis}},\ }\href {\doibase 10.1103/PhysRevD.102.123525} {\bibfield
  {journal} {\bibinfo  {journal} {Phys. Rev. D}\ }\textbf {\bibinfo {volume}
  {102}},\ \bibinfo {pages} {123525} (\bibinfo {year} {2020})},\ \Eprint
  {http://arxiv.org/abs/2005.04115} {arXiv:2005.04115 [gr-qc]} \BibitemShut
  {NoStop}%
\bibitem [{\citenamefont {Luciano}\ and\ \citenamefont
  {Saridakis}(2026)}]{Luciano:2026ufu}%
  \BibitemOpen
  \bibfield  {author} {\bibinfo {author} {\bibfnamefont {G.~G.}\ \bibnamefont
  {Luciano}}\ and\ \bibinfo {author} {\bibfnamefont {E.~N.}\ \bibnamefont
  {Saridakis}},\ }\href@noop {} {\  (\bibinfo {year} {2026})},\ \Eprint
  {http://arxiv.org/abs/2602.20004} {arXiv:2602.20004 [gr-qc]} \BibitemShut
  {NoStop}%
\bibitem [{\citenamefont {Wald}(1984)}]{Wald:1984rg}%
  \BibitemOpen
  \bibfield  {author} {\bibinfo {author} {\bibfnamefont {R.~M.}\ \bibnamefont
  {Wald}},\ }\href {\doibase 10.7208/chicago/9780226870373.001.0001} {\emph
  {\bibinfo {title} {{General Relativity}}}}\ (\bibinfo  {publisher} {Chicago
  Univ. Pr.},\ \bibinfo {address} {Chicago, USA},\ \bibinfo {year}
  {1984})\BibitemShut {NoStop}%
\bibitem [{\citenamefont {Poisson}(2004)}]{poisson2004relativist}%
  \BibitemOpen
  \bibfield  {author} {\bibinfo {author} {\bibfnamefont {E.}~\bibnamefont
  {Poisson}},\ }\href@noop {} {\emph {\bibinfo {title} {A relativist's toolkit:
  the mathematics of black-hole mechanics}}}\ (\bibinfo  {publisher} {Cambridge
  university press},\ \bibinfo {year} {2004})\BibitemShut {NoStop}%
\bibitem [{\citenamefont {Weinberg}(2013)}]{weinberg2013gravitation}%
  \BibitemOpen
  \bibfield  {author} {\bibinfo {author} {\bibfnamefont {S.}~\bibnamefont
  {Weinberg}},\ }\href@noop {} {\emph {\bibinfo {title} {Gravitation and
  cosmology: principles and applications of the general theory of
  relativity}}}\ (\bibinfo  {publisher} {John Wiley \& Sons},\ \bibinfo {year}
  {2013})\BibitemShut {NoStop}%
\bibitem [{\citenamefont {Anand}\ \emph
  {et~al.}(2026{\natexlab{a}})\citenamefont {Anand}, \citenamefont {Devdutt},
  \citenamefont {Jusufi},\ and\ \citenamefont {Saridakis}}]{Anand:2025rjg}%
  \BibitemOpen
  \bibfield  {author} {\bibinfo {author} {\bibfnamefont {A.}~\bibnamefont
  {Anand}}, \bibinfo {author} {\bibfnamefont {S.}~\bibnamefont {Devdutt}},
  \bibinfo {author} {\bibfnamefont {K.}~\bibnamefont {Jusufi}}, \ and\ \bibinfo
  {author} {\bibfnamefont {E.~N.}\ \bibnamefont {Saridakis}},\ }\href {\doibase
  10.1140/epjc/s10052-026-15768-z} {\bibfield  {journal} {\bibinfo  {journal}
  {Eur. Phys. J. C}\ }\textbf {\bibinfo {volume} {86}},\ \bibinfo {pages} {534}
  (\bibinfo {year} {2026}{\natexlab{a}})},\ \Eprint
  {http://arxiv.org/abs/2511.04613} {arXiv:2511.04613 [gr-qc]} \BibitemShut
  {NoStop}%
\bibitem [{\citenamefont {Anand}\ \emph
  {et~al.}(2026{\natexlab{b}})\citenamefont {Anand}, \citenamefont {Jusufi},
  \citenamefont {Basilakos},\ and\ \citenamefont {Saridakis}}]{Anand:2025cer}%
  \BibitemOpen
  \bibfield  {author} {\bibinfo {author} {\bibfnamefont {A.}~\bibnamefont
  {Anand}}, \bibinfo {author} {\bibfnamefont {K.}~\bibnamefont {Jusufi}},
  \bibinfo {author} {\bibfnamefont {S.}~\bibnamefont {Basilakos}}, \ and\
  \bibinfo {author} {\bibfnamefont {E.~N.}\ \bibnamefont {Saridakis}},\ }\href
  {\doibase 10.1140/epjc/s10052-026-15365-0} {\bibfield  {journal} {\bibinfo
  {journal} {Eur. Phys. J. C}\ }\textbf {\bibinfo {volume} {86}},\ \bibinfo
  {pages} {126} (\bibinfo {year} {2026}{\natexlab{b}})},\ \Eprint
  {http://arxiv.org/abs/2512.13769} {arXiv:2512.13769 [gr-qc]} \BibitemShut
  {NoStop}%
\end{thebibliography}%


\end{document}